# Updating Standard Solar Models

F. Ciacio[1],[3] S. Degl'Innocenti[2],[3] and B. Ricci[3]

[1] Dipartimento di Fisica dell'Università di Ferrara, via Paradiso 12, I-44100 Ferrara, Italy
[2] Max-Planck Institut for Astrophysics, K. Schwarzschild Str. 1, D-85740 Garching bei München, Germany
[3] Istituto Nazionale di Fisica Nucleare, Sezione di Ferrara, via Paradiso 12, I-44100 Ferrara, Italy

**Abstract.** We present an updated version of our standard solar model (SSM) where helium and heavy elements diffusion is included and the improved OPAL equation of state (Rogers 1994, Rogers Swenson & Iglesias 1996) is used. In such a way the EOS is consistent with the adopted opacity tables, from the same Livermore group, an occurrence which should further enhance the reliability of the model. The results for the physical characteristics and the neutrino production of our SSM are discussed and compared with previous works on the matter.

In the last decades evolutionary computations of Standard Solar Models (SSM) played a fundamental role in understanding the inner solar structure and, in particular, in approaching the well known problem of solar neutrinos. Bahcall & Pinsonneault (1995; hereafter BP95) have already shown that to reach the agreement with observational evidence given by helioseismology one needs, in addition to the best available input physics, an accurate treatment of element diffusion all along the solar structure. The recent availability of an improved equation of state (EOS), as given by the OPAL group (Rogers 1994, Rogers Swenson & Iglesias 1996) obviously suggests to investigate the influence of such an improvement on SSM. In this paper we discuss this scenario, presenting an updated SSM resulting from a recent version of FRANEC (Frascati Raphson Newton Evolutionary Code), where helium and heavy elements diffusion is included and the OPAL equation of state is used. Note that in such a way the EOS is consistent with the adopted opacity tables, from the same Livermore group (Rogers & Iglesias 1995), an occurrence which should further enhance the reliability of the model.

In addition, updated values of the relevant nuclear cross sections are used and more refined values of the solar constant and age (BP95) are adopted.

Before entering into the argument, let us recall that in recent years helioseismology has added important pieces of information on the solar structure, producing severe tests for standard solar model calculations. According to Christensen-Dalsgaard et al. (1993) one can accurately determine the depth of the solar convective zone and the speed of sound at its bottom:

$$R_b/R_\odot = 0.710 \div 0.716 \qquad (1)$$
$$c_b = 0.221 \div 0.225 \ (Mm/s)$$

Richard et al. (1996; hereafter RCVD96) recently confirmed the value of $R_b$, finding $R_b/R_\odot = 0.7137$. In addition several determinations of the helium photospheric abundances have been derived from inversion of helioseismological data, with results in the range (see Castellani et al. 1996 and refs. therein):

$$Y_{photo} = 0.233 \div 0.268. \qquad (2)$$

The much smaller errors often quoted, reflect the observational frequency errors only. The results actually depend on the method of inversion and on the starting physical inputs (e.g. the EOS), see RCVD96.

Note that, since in building a SSM one is dealing with three free parameters (mixing length, original He content and original Z/X), with these three additional constraints ($R_b$, $c_b$ and $Y_{photo}$) no free parameter is left for SSM builders.

After discussing the effect of the physical inputs, we compare our SSM with other recent solar model calculations, all including diffusion of helium and heavy elements, finding an excellent agreement. We present neutrino fluxes and the expected signals in ongoing experiments. A detailed presentation of our new model tion...) is available on World Wide Web at the address http://dns.unife.it/Fiorentini/index.htm and a more extensive discussion will be published elsewhere.

As for the computations, FRANEC has been described in previous papers (see e.g. Chieffi & Straniero 1989, Castellani et al. 1992). Recent determinations of the the solar luminosity ($L_\odot = 3.844 \cdot 10^{33}$ erg/s) (BP95) and of the solar age ($t_\odot = 4.57 \cdot 10^9$ yr) (BP95) are used. The present ratio of the solar metallicity to solar hydrogen abundance by mass corresponds to the most recent value by Grevesse & Noels (1993) : $(Z/X)_{photo}= 0.0245$.

Following the standard procedure (see e.g. Bahcall & Ulrich 1988), for each set of assumed physical inputs the initial Y, Z and the mixing length parameter $\alpha$ were varied until the radius, luminosity and Z/X at the solar age matched the observed values within a tenth of percent or better. We considered the following steps:

a) As a starting model we used the Straniero (1988) equation of state, the version of the OPAL opacity tables available in 1993 (Rogers & Iglesias 1992, Iglesias Rogers & Wilson 1992) for the Grevesse & Noels (1993) solar metallicity ratio, combined with the molecular opacities by Alexander & Fergusson (1994); diffusion was ignored. This model is useful for a comparison with BP95 (without diffusion) which uses the same chemical composition.

b) Next, we introduced the OPAL equation of state. With respect to other commonly used EOS, this one avoids an ad hoc treatment of the pressure ionization and it provides a systematic expansion in the Coulomb coupling parameter that includes various quantum effects generally not included in other computations (see Rogers 1994, Rogers Swenson & Iglesias 1996 for more details).

c) Next, OPAL 1993 tables were substituted with the latest OPAL opacity tables (Rogers & Iglesias 1995), again for Grevesse & Noels (1993) solar metallicity ratio. With respect to Rogers & Iglesias (1992) the new OPAL tables include the effects on the opacity of seven additional elements and some minor physics changes; moreover the temperature grid has been made denser.

d) Furthermore, we included the diffusion of helium and heavy elements. The diffusion coefficients have been calculated using the subroutine developed by Thoul (see Thoul, Bahcall and Loeb 1994). The variations of the abundance of H, He , C, N, O and Fe are followed all along the solar structure; all these elements are treated as fully ionized. According to Thoul, Bahcall and Loeb (1994) all other elements are assumed to diffuse at the same rate as the fully-ionized iron. To account for the effect of heavy element diffusion on the opacity coefficients we interpolated with different total metallicity (Z=0.01,0.02,0.03,0.04).

e) As a final point, we investigated the effect of updating the nuclear cross sections for $^3$He+$^3$He and $^3$He+$^4$He reactions, following a recent reanalysis of all available data (see Castellani et al. 1996, and table 6). For $S(0)_{34}$, a polynomial fit gives (energy in Mev, $S$ in Mev barn):

$$S_{34}(E) = (4.8 - 2.9E + 0.9E^2) \times 10^{-4}. \quad (3)$$

Alternatively, by using an exponential parametrization, as frequently adopted in the literature, one obtains an equally good fit to experimental data:

$$S_{34}(E) = 5.1 \cdot 10^{-4} exp(-0.83E + 0.25E^2). \quad (4)$$

At the energies of interest, $E_o \approx 20$ Kev, the second expression yields an S factor larger by about 5%, a value which is indicative of the uncertainty on $S_{34}$ at these low energies. In the calculations, we used eq. 4.

The resulting solar models are summarized in table 1, which deserves the following comments:

(a $\to$ b): the introduction of the new OPAL EOS reduces appreciably the initial helium abundance. The Straniero (1988) EOS underestimates the Coulomb effects neglecting the contribution due to the electrons, which are considered as completely degenerate, whereas the OPAL EOS includes corrections for Coulomb forces which are correctly treated (see Rogers 1994 for more details). The models with Straniero EOS have a higher central pressure and a higher central temperature, and correspondingly a higher initial helium abundance. The effect of an underestimated Coulomb correction was discussed in Turck-Chièze & Lopez (1993). Note that the helium abundance is higher than the helioseismological determination by more than $2\sigma$. Moreover, the transition between radiative and convective regions is not correctly predicted by the model, the convective region being definitely too shallow.

(b $\to$ c): the updating of the radiative opacity coefficients has minor effects. The surface helium abundance is now within $2\sigma$ from the helioseismological determination, but the convective zone is again too shallow.

(c $\to$ d): this step shows the effects of diffusion. Helium and heavy elements sink relative to hydrogen in the radiative interior of the star because of the combined effect of gravitational settling and of thermal diffusion. This increases the molecular weight in the core and thus the central temperature is raised. The surface abundances of hydrogen, helium and heavy elements are appreciably affected by diffusion: the initial value $Y_{in}= 0.269$ is reduced to the present photospheric value $Y_{photo}=0.238$, within $2\sigma$ from the helioseismological results. More important, speed are now in good agreement with helioseismological values. With respect to models without diffusion, in the external regions the present helium fraction is reduced while the metal fraction stays at the observed photospheric value. Thus the opacity increases and convection starts deeper in the Sun. As an example of the diffusion process we show in fig.1 the time dependence of the He profile and in fig. 2 the present H profile, calculated with and without diffusion.

(d → e): The modifications of our Solar Model arising from the new values of the nuclear cross sections are negligible with respect to the other improvements just presented.

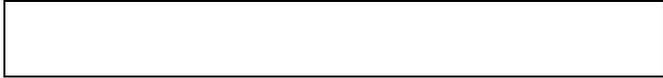

**Fig. 1.** Time dependence of the He profile of our model e). The He profile at t=4.57 Gyr is that of our "best" standard solar model

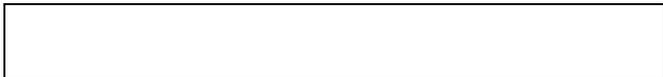

**Fig. 2.** H profile for a solar model without diffusion (our model c) and with diffusion (our model e).

Our "best" Standard Solar Model, model (e), appears in good agreement with recent calculations (table 3) by several authors, all including microscopic diffusion and using (slightly) different physical and chemical inputs, summarized in table 4. Our ingredients are very close to those of BP95 and one notes a substantial similarity with that model. We only have a slightly lower central temperature, possibly an effect of the different EOS. Among the different models, diffusion looks less efficient in RVCD96, as a consequence of the inclusion of rotational induced mixing.

The predicted neutrino fluxes and signals from our SSM are summarized in table 2. All in all, the results are quite stable with respect to the changements we have introduced as long as diffusion is neglected. On the other hand, due to the higher central temperature, model (d) has significantly higher $^8$B and CNO neutrino fluxes. It is essentially the increase of the boron flux which enhances the predicted Chlorine signal. The slight changement in the nuclear cross section weakly affects neutrino fluxes and signals: the $^7$Be and $^8$B fluxes are reduced by about 5%, $S_{34}$. Should we use the polinomial parametrization of eq. 3 one would get a further 5% decrease. As a final remark, we want to stress that again our results are in excellent agreement with other recent calculations including diffusion (see table 5), but the prediction of Kovetz & Shaviv (1994) for the $^{13}$N and $^{15}$O fluxes which are an order of magnitude smaller than those of all other calculations for reasons which are unknown to us [1].

Regarding the comparison with BP95, the small difference in the neutrino production reflects well the small difference in the central temperature, already remarked, and in the adopted values of $S_{34}$ (see table 6).

## 1. Acknowledgments

We warmly thank V. Castellani and G. Fiorentini for their advice. F. Ciacio is grateful to CNR for a fellowship. S. Degl'Innocenti and B. Ricci acknowledge the support of the European Union, through the Human Capital & Mobility Program.

## References

Alexander, D.R. & Ferguson, J.W., 1994, ApJ 437, 879

Bahcall J.N. & Pinsonneault M.H., 1995, Rev. Mod. Phys. 67, 781 (BP95)

Bahcall J.N. & Pinsonneault M.H. 1992, Rev. Mod. Phys. 64, 885 (BP92)

Bahcall J.N. & Ulrich R.K., 1988, Rev. Mod. Phys. 60, 297

Castellani V., Chieffi A., Straniero O., 1992, ApJS 78, 517

Castellani V., Degl'Innocenti S., Fiorentini G., Lissia M. and Ricci B., preprint INFNFE-10-96, submitted to Phys. Rep. 1996.

Caughlan G. R. & Fowler W. A., 1988, Atomic Data Nucl. data tables 40, 283 (1988)

Chieffi A. & Straniero O., 1989, ApJS 71, 47

Christensen-Dalsgaard J., Gough D. O. and Thompson M. J. 1991 ApJ 378, 413

Christensen-Dalsgaard J. & Dappen W. 1992, A&A Rev. 4, 267

Christensen-Dalsgaard J., Proffit C. R., and Thompson M. J., 1993, ApJL, 403, L75

Cox A. N. & Steward J. N., 1970, ApJS 19, 243

Cox, A. J., Guzik A. and Kidman P. B., 1989, ApJ 342, 1187

Dappen W., Gough D. O. and Thompson M. J. 1988 in "Seismology of the Sun and Sun-like stars" E.J. Rolfe ed., ESTEC, Noordwijk

Dappen W., Gough D. O., A. G. Kosovichev and Thompson M. J 1991 in "Challenges to the theories of the structure of moderate-mass stars", Gough D. O. and Toomre J. eds., Springer, Heidelberg

Dar A. & Shaviv G. 1996, preprint e-archive astro-ph/9604009

---

[1] After the completion of this paper we received a preprint by Dar & Shaviv (1996) where the calculated $^{13}$N ($^{15}$O) neutrino fluxes are now 0.382 (0.374)·$10^9$ cm$^{-2}$s$^{-1}$, much closer to the general predictions.

1991, Mon. Not. R. Astr. Soc., 249, 602 (1991)

Dziembowski W. A., Goode P.A., Pamyatnykh A. A and Sienkiewicz A. 1994, ApJ 432, 417

Eggleton P.P., Faulkner J. & Flannery B. P. 1973, A&A 23, 235

Fowler W.A. , Caughlan G. R. and Zimmerman B. A. 1975 Ann. Rev. A&A 13, 113

Hernandez F.P. & Christensen-Dalsgaard J., 1994, Mon. Not. R. Astron. Soc. 269, 475

Grevesse N. & Noels A., 1993a, in "Origin and Evolution of the elements", ed. N. Prantzos, E. Vangioni-Flam, M. Casse (Cambridge Univ. Press, Cambridge), p.15

Grevesse N. 1991, in "Evolution of stars: the photospheric abundance connection", ed. Michaud G. and Tutukov A. IAU p. 63

Iglesias C. A., Rogers F.J. and Wilson B.G. 1992, ApJ 397,717

Kovetz A. & Shaviv G. 1994, ApJ 426, 787

Mihalas D., Dappen W. & Hummer D. G. 1988, ApJ 331, 815

Noerdlinger P. D. 1977, A&A 57,507

Proffit C. R., 1994, ApJ 425, 849

Rogers, F.J., Iglesias, C.A., 1992, ApJS 79, 507

Rogers, F.J., 1994, in IUA Colloq. 147, "The Equation of State in Astrophysics", ed. Chabrier, G., Schatzman,E.L. Cambridge University Press, Cambridge, p. 16

Rogers, F.J., Iglesias, C.A., 1995, in ASP Conference series "Astrophysical application of powerful new database" ed. Adelman, S.J., Wiesse, W.L., vol. 78, p. 31

Richard O., Vauclair S., Charbonnel C. and Dziembowski W. A. 1996, submitted to A&A (RCVD96)

Rogers, F.J., Swenson, F.J. and Iglesias, C.A., 1996, ApJ 456, 902

Turck-Chièze S. & Lopez I. , 1993, ApJ 408, 367

Thoul, A. A., Bahcall J. N. and Loeb A. 1994, ApJ 421, 828

Straniero, O., 1988, A&AS 76, 157

Vorontov S. V., Baturin V. A. and Pamyatnykh A. A.,1991, Nature 349, 49

Table 1: Comparison among solar models obtained with different versions of the FRANEC code. The labels (a) to (e) correspond to the models defined in the text. Our best Standard Solar Model is (e). The last column shows the helioseismological results as discussed in the text. Here: S88=Straniero (1988)

|  | (a) | (b) | (c) | (d) | (e) best | Helioseism. |
|---|---|---|---|---|---|---|
| $t_\odot$[Gyr] | 4.57 | 4.57 | 4.57 | 4.57 | 4.57 |  |
| $L_\odot[10^{33}$ erg/cm$^2$/s] | 3.846 | 3.843 | 3.844 | 3.843 | 3.844 |  |
| $R_\odot[10^{10}$ cm] | 6.961 | 6.963 | 6.959 | 6.959 | 6.960 |  |
| $(Z/X)_{photo}$ | 0.0245 | 0.0245 | 0.0245 | 0.0245 | 0.0245 |  |
| $\alpha$ | 2.023 | 1.774 | 1.786 | 1.904 | 1.901 |  |
| $X_{in}$ | 0.699 | 0.718 | 0.722 | 0.711 | 0.711 |  |
| $Y_{in}$ | 0.284 | 0.265 | 0.261 | 0.269 | 0.269 |  |
| $Z_{in}$ | 0.0171 | 0.0176 | 0.0177 | 0.0198 | 0.0198 |  |
| $X_{photo}$ | 0.699 | 0.718 | 0.722 | 0.743 | 0.744 |  |
| $Y_{photo}$ | 0.284 | 0.265 | 0.261 | 0.238 | 0.238 | $0.233 \div 0.268$ |
| $Z_{photo}$ | 0.0171 | 0.0176 | 0.0177 | 0.0182 | 0.0182 |  |
| $R_b/R_\odot$ | 0.738 | 0.726 | 0.728 | 0.716 | 0.716 | $0.710 \div 0.716$ |
| $T_b[10^6$ K] | 1.99 | 2.10 | 2.08 | 2.17 | 2.17 |  |
| $c_b[10^7$ cm s$^{-1}$] | 2.11 | 2.16 | 2.16 | 2.22 | 2.22 | $2.21 \div 2.25$ |
| $T_c[10^7$ K] | 1.555 | 1.545 | 1.542 | 1.569 | 1.569 |  |
| $\rho_c$[100 gr cm$^{-3}$] | 1.524 | 1.472 | 1.470 | 1.514 | 1.518 |  |
| $Y_c$ | 0.63 | 0.61 | 0.61 | 0.63 | 0.63 |  |
| OPACITY | OPAL | OPAL | OPAL | OPAL | OPAL |  |
| EOS | S88 | OPAL | OPAL | OPAL | OPAL |  |

Table 2: Neutrino fluxes and signals obtained with different versions of the FRANEC code. The labels (a) to (e) correspond to the models defined in the text. Our best prediction is (e).

|  | (a) | (b) | (c) | (d) | (e) best |
| --- | --- | --- | --- | --- | --- |
| $\Phi_{pp}$ [$10^9$ cm$^{-2}$ s$^{-1}$] | 60.17 | 60.37 | 60.66 | 59.76 | 59.92 |
| $\Phi_{pep}$ [$10^9$ cm$^{-2}$ s$^{-1}$] | 0.14 | 0.14 | 0.14 | 0.14 | 0.14 |
| $\Phi_{Be}$ [$10^9$ cm$^{-2}$ s$^{-1}$] | 4.58 | 4.22 | 4.09 | 4.71 | 4.49 |
| $\Phi_{N}$ [$10^9$ cm$^{-2}$ s$^{-1}$] | 0.39 | 0.36 | 0.35 | 0.52 | 0.53 |
| $\Phi_{O}$ [$10^9$ cm$^{-2}$ s$^{-1}$] | 0.33 | 0.30 | 0.29 | 0.45 | 0.45 |
| $\Phi_{B}$ [$10^6$ cm$^{-2}$ s$^{-1}$] | 4.73 | 4.18 | 3.95 | 5.37 | 5.16 |
| $S_{Ga}$ [SNU] | 126 | 121 | 120 | 130 | 128 |
| $S_{Cl}$ [SNU] | 6.9 | 6.2 | 5.9 | 7.7 | 7.4 |

Table 3: Comparison among several SSMs, all including diffusion. The correspondence between acronyms and references is as follows: CGK89=Cox et al. (1989), P94=Proffit (1994), KS94=Kovetz & Shaviv (1994), RVCD96= Richard et al. (1996). BP95 indicates here the "best model with diffusion" of Bahcall & Pinsonneault 1995. FRANEC96 indicates our best model with diffusion, model (e). The star indicates values of $c_b$ calculated by us assuming fully ionized gas EOS.

|  | CGK89 | P94 | KS94 | RVCD96 | BP95 | FRANEC96 |
| --- | --- | --- | --- | --- | --- | --- |
| $t_\odot$ [Gyr] | 4.54 | 4.60 | 4.54 | 4.60 | 4.57 | 4.57 |
| $L_\odot$ [$10^{33}$ erg/cm$^2$/s] | 3.828 | 3.846 | 3.8515 | 3.851 | 3.844 | 3.844 |
| $R_\odot$ [$10^{10}$ cm] | 6.9599 | 6.9599 | 6.960 | 6.959 | 6.9599 | 6.960 |
| $(Z/X)_{photo}$ | 0.02464 | 0.02694 | 0.02763 | 0.0263 | 0.02446 | 0.0245 |
| $X_{in}$ | 0.691 | 0.6984 | 0.6797 | 0.7012 | 0.70247 | 0.711 |
| $Y_{in}$ | 0.289 | 0.2803 | 0.2991 | 0.2793 | 0.27753 | 0.269 |
| $Z_{in}$ | 0.02 | 0.02127 | 0.0211 | 0.0195 | 0.02 | 0.0198 |
| $X_{photo}$ | 0.7265 | 0.7290 | 0.7050 | 0.7226 | 0.73507 | 0.744 |
| $Y_{photo}$ | 0.2556 | 0.2514 | 0.2754 | 0.2584 | 0.24695 | 0.238 |
| $Z_{photo}$ | 0.0179 | 0.01964 | 0.01948 | 0.0190 | 0.01798 | 0.0182 |
| $R_b/R_\odot$ | 0.721 | 0.7115 | 0.7167 | 0.716 | 0.712 | 0.716 |
| $T_b$ [$10^6$ K] | 2.142 |  | 2.151 | 2.175 | 2.204 | 2.17 |
| $c_b$ [$10^7$ cm s$^{-1}$] | 2.21* |  | 2.19* | 2.22* | 2.25* | 2.22 |
| $T_c$ [$10^7$ K] | 1.573 | 1.581 | 1.567 | 1.567 | 1.5843 | 1.569 |
| $\rho_c$ [100 gr cm$^{-3}$] | 1.633 | 1.559 | 1.550 | 1.545 | 1.562 | 1.518 |

Table 4: Physical inputs of the SSMs in table 3: C&S70=Cox & Steward (1970), MHD= Mihalas et al. (1988), CEFF=EOS of Eggleton et al. (1973) with the Coulombian correction added (see Christensen -Dalsgaard & Dappen 1992), EFF=Eggleton et al. (1973), BP92=Bahcall & Pinsonneault (1992), G&N93= Grevesse & Noels (1993), G91=Grevesse (1991), F75=Fowler et al. (1975), C&F88=Caughlan et al. (1988).

|                | GGK89  | P94  | KS94  | RVCD96 | BP95    | FRANEC96 |
|----------------|--------|------|-------|--------|---------|----------|
| OPACITY        | C&S70  | OPAL | OPAL  | OPAL   | OPAL    | OPAL     |
| EOS            | MHD    | CEFF | KS94  | MHD    | BP92    | OPAL     |
| MIXTURE        | G&N93  | G91  | G&N93 | G&N93  | G&N93   | G&N93    |
| CROSS SECTIONS | F75    | BP92 | C&F88 | C&F88  | table 6 | table 6  |

Table 5: Comparison among the neutrino fluxes and signals of the SSMs in table 3.

|                                              | P94   | KS94   | RVCD96 | BP95  | FRANEC96 |
|----------------------------------------------|-------|--------|--------|-------|----------|
| $\Phi_{pp}$ [$10^9$ cm$^{-2}$ s$^{-1}$]      | 59.1  | 59.9   | 59.4   | 59.1  | 59.92    |
| $\Phi_{pep}$ [$10^9$ cm$^{-2}$ s$^{-1}$]     | 0.139 | 0.138  | 0.138  | 0.140 | 0.14     |
| $\Phi_{Be}$ [$10^9$ cm$^{-2}$ s$^{-1}$]      | 5.18  | 4.91   | 4.8    | 5.15  | 4.49     |
| $\Phi_{N}$ [$10^9$ cm$^{-2}$ s$^{-1}$]       | 0.64  | 0.0611 | 0.559  | 0.618 | 0.53     |
| $\Phi_{O}$ [$10^9$ cm$^{-2}$ s$^{-1}$]       | 0.557 | 0.0178 | 0.481  | 0.545 | 0.45     |
| $\Phi_{B}$ [$10^6$ cm$^{-2}$ s$^{-1}$]       | 6.48  | 5.83   | 6.33   | 6.62  | 5.16     |
| $S_{Ga}$[SNU]                                | 136.9 | 124.3  | 132.77 | 137.0 | 128      |
| $S_{Cl}$[SNU]                                | 9.02  | 7.9    | 8.49   | 9.3   | 7.4      |

Table 6: Astrophysical $S$-factors [MeV barn] and their derivatives with respect to energies $S'$ [barn] for our "best" solar model, for the "best SSM with diffusion" of Bahcall & Pinsonneault 1995 (BP95) and for our models.

|  | BP95 | models (a-d) | model (e) (our best) |
|---|---|---|---|
| $S(0)_{pp}$ | $3.89 \times 10^{-25}$ | $3.89 \times 10^{-25}$ | $3.89 \times 10^{-25}$ |
| $S'(0)_{pp}$ | $4.52 \times 10^{-24}$ | $4.52 \times 10^{-24}$ | $4.52 \times 10^{-24}$ |
| $S(0)_{33}$ | 4.99 | 5.00 | 5.1 |
| $S'(0)_{33}$ | -0.9 | -0.9 | -3.0 |
| $S(0)_{34}$ | $5.24 \times 10^{-4}$ | $5.33 \times 10^{-4}$ | $5.1 \times 10^{-4}$ |
| $S'(0)_{34}$ | $-3.1 \times 10^{-4}$ | $-3.10 \times 10^{-4}$ | $-4.23 \times 10^{-4}$ |
| $S(0)_{17}$ | $2.24 \times 10^{-5}$ | $2.24 \times 10^{-5}$ | $2.24 \times 10^{-5}$ |
| $S'(0)_{17}$ | $-3.00 \times 10^{-5}$ | $-3.00 \times 10^{-5}$ | $-3.00 \times 10^{-5}$ |
| $S(0)_{^{12}C+p}$ | $1.45 \times 10^{-3}$ | $1.40 \times 10^{-3}$ | $1.40 \times 10^{-3}$ |
| $S'(0)_{^{12}C+p}$ | $2.45 \times 10^{-4}$ | $4.24 \times 10^{-3}$ | $4.24 \times 10^{-3}$ |
| $S(0)_{^{13}C+p}$ | $5.50 \times 10^{-3}$ | $5.50 \times 10^{-3}$ | $5.50 \times 10^{-3}$ |
| $S'(0)_{^{13}C+p}$ | $1.34 \times 10^{-2}$ | $1.34 \times 10^{-2}$ | $1.34 \times 10^{-2}$ |
| $S(0)_{^{14}N+p}$ | $3.29 \times 10^{-3}$ | $3.32 \times 10^{-3}$ | $3.32 \times 10^{-3}$ |
| $S'(0)_{^{14}N+p}$ | $-5.91 \times 10^{-3}$ | $-5.91 \times 10^{-3}$ | $-5.91 \times 10^{-3}$ |
| $S(0)_{^{15}N(p,\gamma)^{16}O}$ | $6.40 \times 10^{-2}$ | $6.40 \times 10^{-2}$ | $6.40 \times 10^{-2}$ |
| $S'(0)_{^{15}N(p,\gamma)^{16}O}$ | $3.00 \times 10^{-2}$ | $3.00 \times 10^{-2}$ | $3.00 \times 10^{-2}$ |
| $S(0)_{^{15}N(p,\alpha)^{12}C}$ | $7.80 \times 10$ | $7.04 \times 10$ | $7.04 \times 10$ |
| $S'(0)_{^{15}N(p,\alpha)^{12}C}$ | $3.51 \times 10^{2}$ | $4.21 \times 10^{2}$ | $4.21 \times 10^{2}$ |
| $S(0)_{^{16}O+p}$ | $9.40 \times 10^{-3}$ | $9.40 \times 10^{-3}$ | $9.40 \times 10^{-3}$ |
| $S'(0)_{^{16}O+p}$ | $-2.30 \times 10^{-2}$ | $-2.30 \times 10^{-2}$ | $-2.30 \times 10^{-2}$ |

Fig. 1

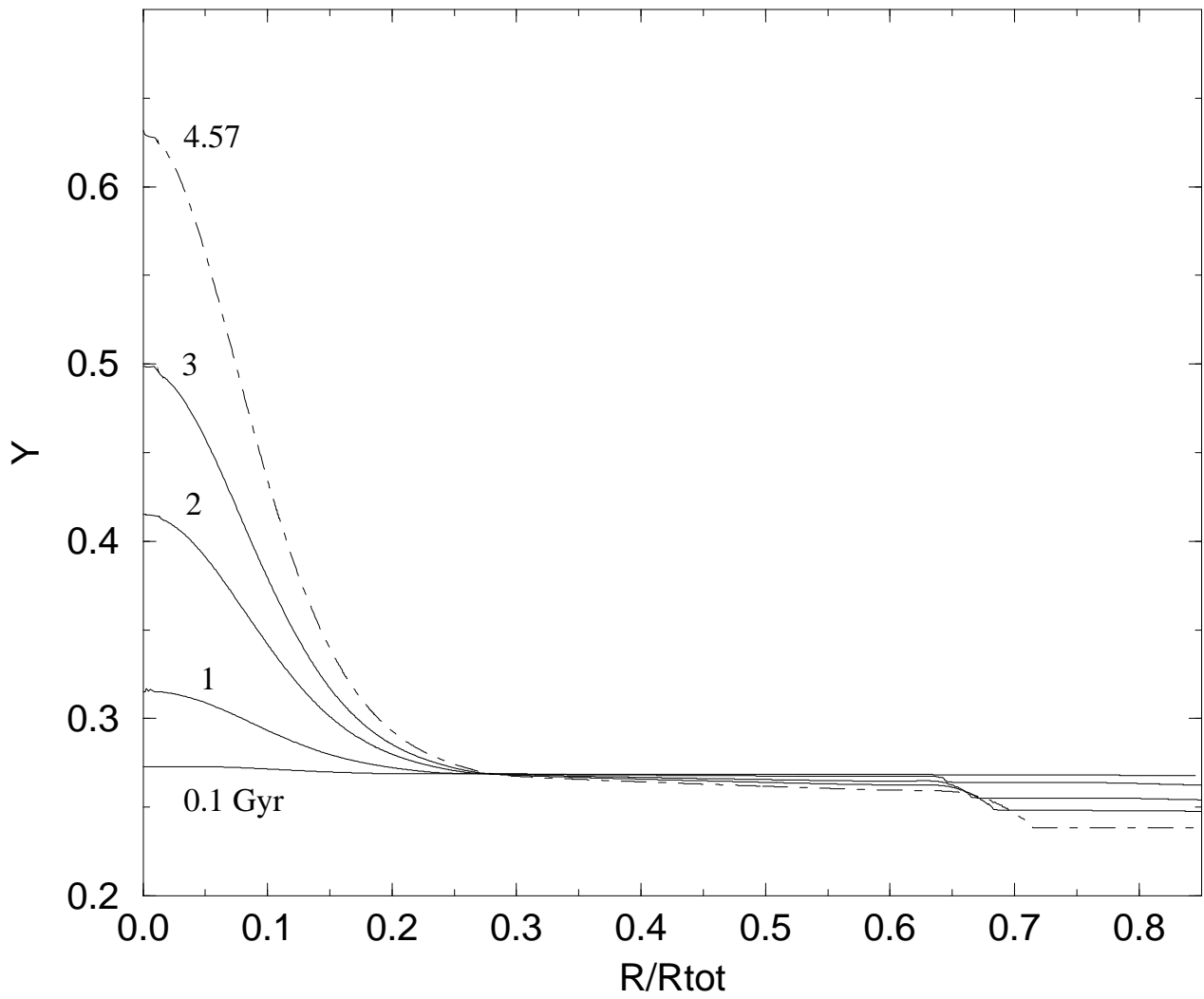

Fig. 2

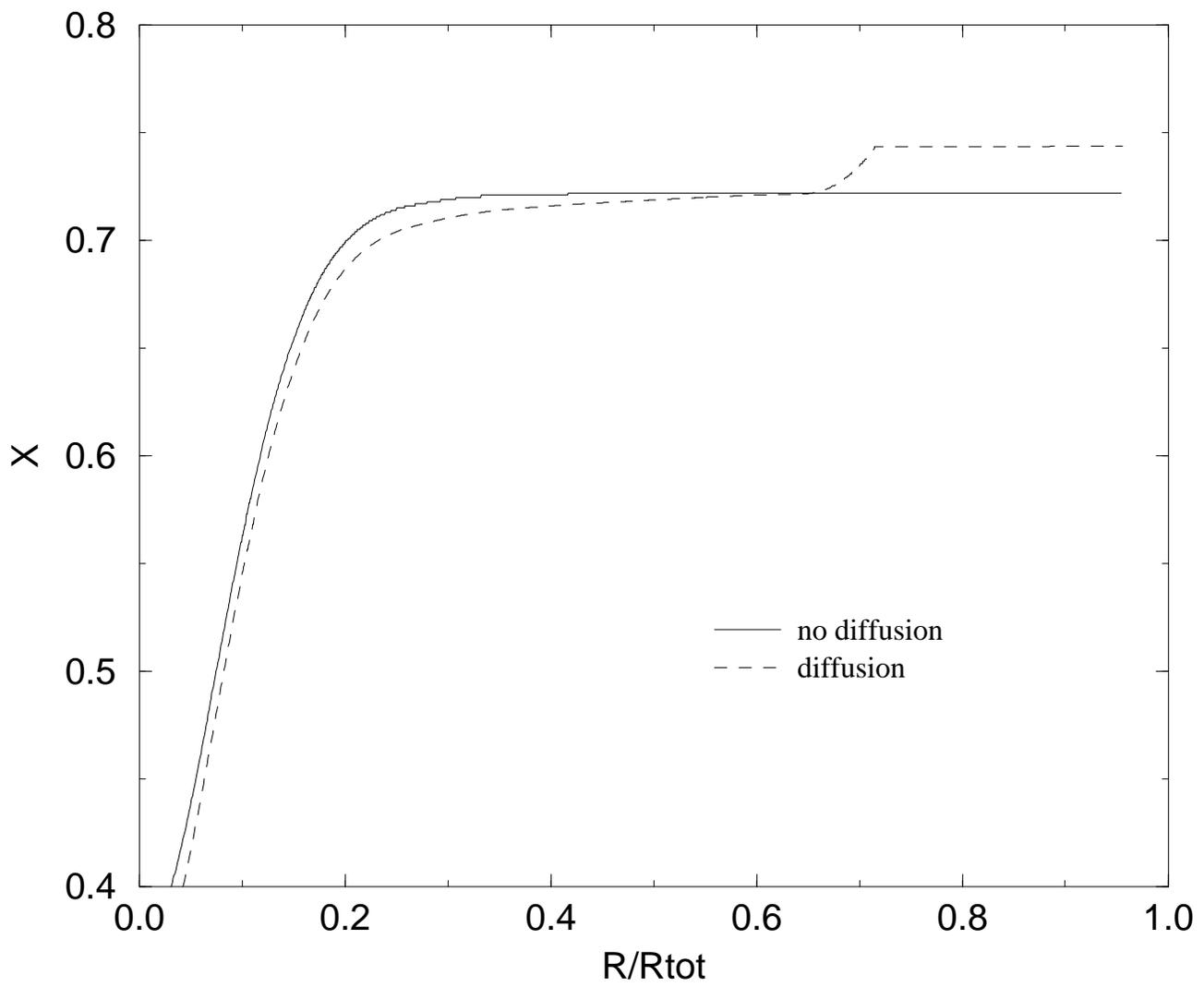